\documentclass[aps,prb,twocolumn,showpacs,superscriptaddress]{revtex4-1}

\usepackage{graphicx}
\usepackage{amsmath}
\usepackage{multirow}
\usepackage{tabularx}
\usepackage{color}
\usepackage[colorlinks,linkcolor=blue,anchorcolor=blue,citecolor=blue]{hyperref}

\begin{document}

\title{Crystal and electronic structure of a quasi-two-dimensional semiconductor Mg$_3$Si$_2$Te$_6$}
\author{Chaoxin Huang}
\email{The authors contributed equally to this work.}
\affiliation{Center for Neutron Science and Technology, Guangdong Provincial Key Laboratory of Magnetoelectric Physics and Devices, School of Physics, Sun Yat-Sen University, Guangzhou, 510275, China }
\author{Benyuan Cheng}
\email{The authors contributed equally to this work.}
\affiliation{Shanghai Institute of Laser Plasma, Shanghai 201800, China}
\affiliation{Center for High Pressure Science and Technology Advanced Research, Shanghai 201203, China}
\author{Yunwei Zhang}
\affiliation{Center for Neutron Science and Technology, Guangdong Provincial Key Laboratory of Magnetoelectric Physics and Devices, School of Physics, Sun Yat-Sen University, Guangzhou, 510275, China }
\author{Long Jiang}
\affiliation{Instrumentation Analysis and Research Center, Sun Yat-Sen UniVersity, Guangzhou 510275, China}
\author{Lisi Li}
\author{Mengwu Huo}
\author{Hui Liu}
\author{Xing Huang}
\author{Feixiang Liang}
\author{Lan Chen}
\author{Hualei Sun}
\author{Meng Wang}
\email{wangmeng5@mail.sysu.edu.cn}
\affiliation{Center for Neutron Science and Technology, Guangdong Provincial Key Laboratory of Magnetoelectric Physics and Devices, School of Physics, Sun Yat-Sen University, Guangzhou, 510275, China }

\begin{abstract}

We report the synthesis and characterization of a Si-based ternary semiconductor Mg$_3$Si$_2$Te$_6$, which exhibits a quasi-two-dimensional structure, where the trigonal Mg$_2$Si$_2$Te$_6$ layers are separated by Mg ions. Ultraviolet-visible absorption spectroscopy and density functional theory calculations were performed to investigate the electronic structure. The experimentally determined direct band gap is 1.39 eV, consistent with the value of the density function theory calculations. Our results reveal that Mg$_3$Si$_2$Te$_6$ is a direct gap semiconductor with a relatively narrow gap, which is a potential candidate for infrared optoelectronic devices.

\end{abstract}
\maketitle

\section{INTRODUCTION}

Narrow-gap semiconductors (NGS) play an important role in optoelectronic and magnetic devices. Because of their unique advantages, the properties and applications of NGS have been widely studied for decades in a variety of fields, such as magnetic field sensors, photovoltaics, infrared photodetectors, lasers and so on\cite{elliott1998,maier1980,stradling1996,xie2021,zhang2017}. Magnetic field sensors, including magnetoresistors and Hall sensors, are generally used in magnetic recording and magnetic measurement technologies\cite{ripka2010,heremans1993}. Materials for magnetic field sensors are doped semiconductors with high electronic density of states and mobility, such as InSb and InAs which are extremely sensitive to magnetic field\cite{solin2000,berus2004,mihajlovic2005}.
Conventional photovoltaic materials mostly use wide-gap semiconductors that utilize photons in the ultraviolet and visible regions, limiting the conversion efficiency. Tuning the band gap is considered to be one of the effective solutions\cite{cheng2020}. NGS-based photoelectrodes can absorb light at longer wavelengths and enhance the solar energy conversion efficiency\cite{zheng2019}.
As a more widely used field, infrared photodetectors, such as the quantum dots-in-a-well photodetectors, have developed rapidly in recent years\cite{downs2013,martyniuk2008,chen2018}. In addition, NGS lasers have attracted more and more attention due to their stability, tunability, simple structure, and many other advantages\cite{tournie2012,harman1974}.

One of the most important parameters for photoelectric materials is the value of the electronic band gap. Hg$_{1-x}$Cd$_x$Te and Pb$_{1-x}$Sn$_x$Te are among the hottest researched NGS systems in recent years, whose gaps can be tuned by compositions in a wide range, even close to zero\cite{harman1974,rogalski2005,lei2015,piotrowski2004}. Alloying is also an effective method to modulate the band gap, which is beneficial for the application of the mid- and long-infrared photodetectors\cite{yang2022}. Although InSb and HgCdTe based detectors have been well available in commercial applications, their poor flexibility, instability of the material interface, complex manufacturing process, and low temperature working environment limit their further development\cite{piotrowski1998,jiao2022,yang2022}. Therefore, it is crucial to search for more NGS that are suitable for manufacturing stable and efficient infrared detectors. Meanwhile, new NGS with excellent performance also have potential applications in laser, photovoltaics, and other magnetic devices.

In this work we report the successful synthesis of a Si-based nonmagnetic semiconductor Mg$_3$Si$_2$Te$_6$, whose structure is found to exhibit a quasi-two-dimensional (quasi-2D) layered configuration. Fittings using a direct and indirect band gap models to the ultraviolet-visible (UV-vis) absorption spectra result in a direct gap of 1.39 eV or an indirect gap of 0.6 eV, respectively. Electronic structure calculations using density functional theory (DFT) yield a direct band gap of 1.2 eV, close to the direct gap fitted from the UV-vis absorption spectra, revealing that Mg$_3$Si$_2$Te$_6$ is a direct gap semiconductor. The band gap is narrower than the energies of visible photons, allowing Mg$_3$Si$_2$Te$_6$ a potential candidate for infrared photoelectric material.

\begin{figure}[b]
\includegraphics[scale=0.33]{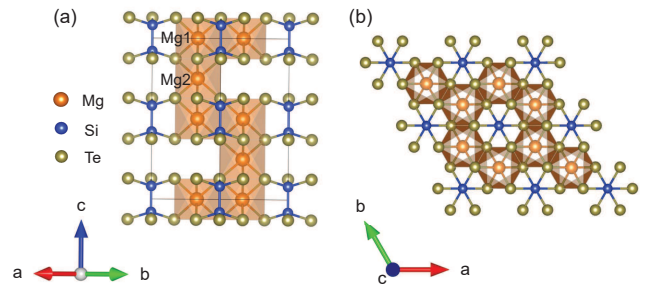}
\caption{Crystal structure of Mg$_3$Si$_2$Te$_6$ (space group: $P\overline{3}1c$) viewed (a) perpendicular to the $c$-axis and (b) parallel to the $c$-axis, respectively. The orange, blue, and grey balls represent Mg, Si, and Te. }
\label{fig1}
\end{figure}

\begin{table}[t]
\caption{Parameters of Mg$_3$Si$_2$Te$_6$ refined from single-crystal XRD at 253K.}
\begin{tabular}{ccccccc}
\hline \hline
Empirical formula      & Mg$_3$Si$_2$Te$_6$   \\ \hline
Formula weight   & 894.69 \\
Temperature   & 252.99(10) K \\
Crystal system   & trigonal \\
Space group & $P\overline{3}1c$    \\
Unit-cell parameters & $a$ = $b$ = 7.0642(9) {\AA}    \\
                     & $c$ = 14.464(2) {\AA}      \\
                     & $\alpha$ = $\beta$ = 90$^\circ$         \\
                     & $\gamma$ = 120$^\circ$             \\
Atomic parameters               \\
Mg1                  & 4f (1/3, 2/3, 0.4987(5))  \\
Mg2                  & 2d (2/3, 4/3, 1/4)    \\
Si                   & 4e (0, 1, 0.4192(4))  \\
Te                   & 12i ($x$, $y$, $z$)  \\
                     & $x$ = 0.3402(1), $y$ = 0.9990(2),  \\
                     & $z$ = 0.3733(1)    \\
Volume      & 625.11(18) {\AA}$^3$   \\
Density     & 4.753 g/cm$^3$              \\
Absorp. coeff.      & 14.102 mm$^{-1}$  \\
$F$(000)    & 725.0          \\
Crystal size       & 0.09 $\times$ 0.04 $\times$ 0.02 mm$^3$   \\
Radiation      & Mo K$_\alpha$ ($\lambda$ = 0.7107 {\AA} )     \\
2$\Theta$ range for data collection  & 5.632$^\circ$ to 53.96$^\circ$  \\
Index ranges  & -9 $\leq$ h $\leq$ 7, -9 $\leq$ k $\leq$ 9, \\ &-18 $\leq$ l $\leq$ 18  \\
Reflections collected  & 3733  \\
Independent reflections  & 439  \\
Data/restraints/parameters  & 439/0/18  \\
Goodness-of-fit on $F$$^2$   & 0.904  \\
Final R indexes [$I$ $\geq$ 2$\sigma$(I)]  & $R$$_1$ = 0.0391,  $wR$$_2$ = 0.0790  \\
Final R indexes [all data]	 & $R$$_1$ = 0.0767, $wR$$_2$ = 0.1014  \\
Largest diff. peak/hole/\emph{e} {\AA}$^{-3}$  & 1.38/ - 0.96   \\   \hline \hline
\end{tabular}
\label{table}
\end{table}

\section{EXPERIMENT AND CALCULATION}

Single crystal samples of Mg$_3$Si$_2$Te$_6$ were grown by a self-flux method\cite{Yin2020,Sun2021,Li2021}. Mg (99.99$\%$) robs, Si (99.99$\%$) powders, and Te blocks (99.99$\%$) were mixed in a molar ratio of $3:2:6$ after rough grinding, then sealed in an evacuated quartz ampoule. The ampoule was firstly heated to 600 $^{\circ}$C in 10 h and held for 15 h, then heated to 1050 $^{\circ}$C in 20 h and dwelled for 10 h, followed by a slow cooling to 750 $^{\circ}$C in 150 h before the furnace was shut down. Shiny plate-like single crystals were obtained. The samples are air sensitive and were stored in an argon-filled glove box to minimize the exposure to air.

\begin{figure}[t]
\includegraphics[scale=0.55]{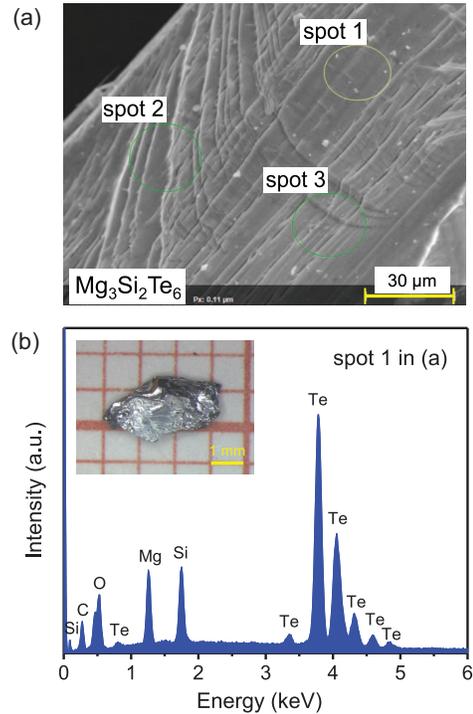}
\caption{ (a) A zoomed-in view on the surface of the Mg$_3$Si$_2$Te$_6$ single crystal taken from SEM. The three circles mark the measured spots. (b) An EDS spectrum of the spot 1 in (a). The elements represented by different characteristic peaks are indicated. Small amounts of carbon and oxygen may derive from the background of conductive adhesive tape. The inset is a photo of the single crystal under an optical microscope. }
\label{fig2}
\end{figure}

Single-crystal x-ray diffraction (XRD) measurements on a single-crystal x-ray diffractometer (SuperNova, Rigaku), a scanning electron microscopy (SEM) photography, and an energy dispersive x-ray spectroscopy (EDS) (EVO, Zeiss) were employed to determine the crystal structure, morphology and composition. Resistivity was measured using the standard four-probe method on single crystals with a typical size of 3 $\times1\times0.5$ mm$^3$ on a physical property measurement system (PPMS) (Quantum Design). UV-vis absorption spectroscopy measurements were conducted using an Ocean Optics DH-2000-BAL spectrometer. The sample was pre-peeled into thin flakes with right thickness and then loaded into a diamond anvil cell (DAC) to avoid oxidation. The outboard ring of the DAC is the T301 steel gasket, which was pre-indented and drilled a hole with a diameter of ~150 $\mu$m to serve as the sample chamber. Each spectrum was collected for 2 seconds with wavelengthes ranged from 180 to 860 nm. The signal from background was obtained by shining the beam on a spot inside the sample chamber but away from the sample.

Electronic band structure was calculated using the DFT within the Perdew-Burke-Ernzerhof (PBE) exchange-correlation as implemented in the Vienna $Ab~initio$ Simulation Package (VASP) code\cite{Perdew1996,Vargas2020}. A plane wave energy cutoff of 600 eV and dense meshes of Monkhorst-Pack $k$-points were used to ensure that all calculations are well converged to 1 meV/atom.

\begin{figure}[t]
\includegraphics[scale=0.26]{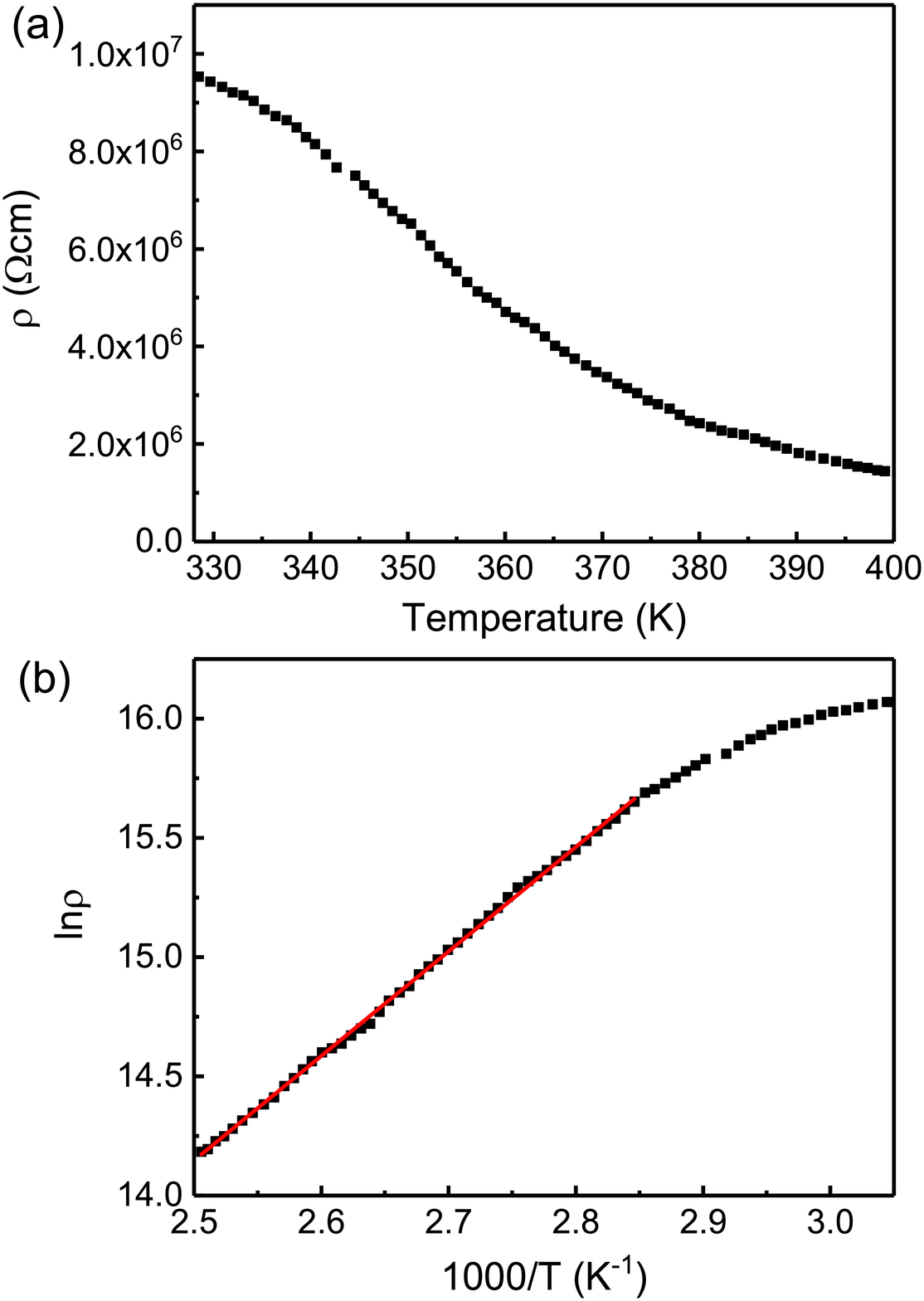}
\caption{(a) Resistivity of Mg$_3$Si$_2$Te$_6$ in the temperature range of 328$\sim$400 K. (b) $\ln\rho$ as a function of $1000/T$. The red line is a fitting using the thermal activation-energy modol $\rho$($T$) = $\rho$$_0$exp($E$$_a$/$k$$_B$$T$). }
\label{fig3}
\end{figure}

\section{RESULTS AND DISCUSSION}

The structure of Mg$_3$Si$_2$Te$_6$ is refined from XRD measurements on single crystals. The structural parameters are summarized in Table \ref{table}. Mg$_3$Si$_2$Te$_6$ crystalizes in the trigonal space group $P\overline{3}1c$ with $a$ = 7.0642(9) {\AA} and $c$ = 14.464(2) {\AA} at 253 K and exhibits a quasi-2D structure. The crystal structures viewed from different directions are illustrated in Fig. \ref{fig1}. The layer of the $ab$ plane constituting the formula as Mg$_2$Si$_2$Te$_6$ consists of edge shared MgTe$_6$ octahedra and Si-Si dimers. The Mg$_2$Si$_2$Te$_6$ layer is isomorphic to a widely studied 2D van der Waals ferromagnetic compound Cr$_2$Si$_2$Te$_6$\cite{carteaux1995,cai2020}. The layers are linked by Mg2 that is half of Mg1, constituting Mg$_3$Si$_2$Te$_6$. Mg$_3$Si$_2$Te$_6$ is isostructural to Mn$_3$Si$_2$Te$_6$\cite{vincent1986} and Mn$_3$Si$_2$Se$_6$\cite{may2020} which are quasi-2D semiconductors with enriched properties such as ferrimagnetism and large colossal magnetoresistance\cite{Ni2021,Seo2021}.

To determine the composition of the samples, we performed EDS measurements on Mg$_3$Si$_2$Te$_6$ single crystals. Figure \ref{fig2}(a) is a zoomed-in view of the measured crystal in the $ab$ plane taken from SEM. It is obvious that the crystal is layered with some Te flux on the surface. By normalizing the content of Mg to be 3, the composition determined from EDS is Mg$_3$Si$_{1.87(4)}$Te$_{6.8(2)}$. The ratio of Mg to Si ratio is close to expectation. The exceeding content of Te of 13\% can be attributed to the excess Te flux on the surface.

\begin{figure}[t]
\includegraphics[scale=0.25]{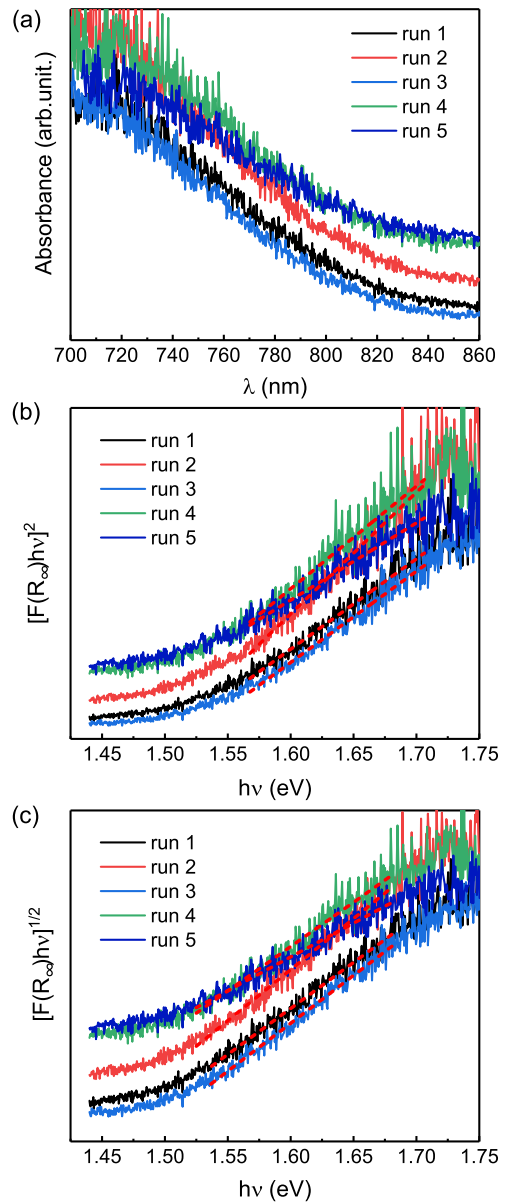}
\caption{(a) UV-vis absorption spectra of Mg$_3$Si$_2$Te$_6$ in a DAC at room temperature.  (b) The Tauc plots of [$F$($R$$_\infty$)$h$$\nu$]$^{1/n}$ = $A$($h$$\nu$ - $E_g$) versus $h\nu$ with $n$ = 1/2 for direct band gap and (c) $n$ = 2 for indirect band gap. The dotted red lines are fittings of the linear regions. }
\label{fig4}
\end{figure}

Figure \ref{fig3} shows the resistivity measured between 328 and 400 K. The resistivity below 328 K exceeding 10$^6$ $\Omega$cm is beyond the measured range of the instrument. As shown in Fig. \ref{fig3} (a), the resistivity exhibits an insulating behavior, and no obvious anomaly can be observed up to 400 K. Figure \ref{fig3} (b) is a plot of the resistivity in $\ln\rho$ as a function of 1000/$T$. A fitting using the thermal activation-energy model $\rho$($T$) = $\rho$$_0$ exp($E$$_a$/$k$$_B$$T$) to the linear segment that corresponds to 350 to 400 K results in $E_a=0.378$ eV. In the model, $\rho$$_0$ is a prefactor, $k$$_B$ is the Boltzmann constant, and $E_a$ is the thermal activation energy. The obtained activation energy of 0.378 eV indicates that the energy gap of Mg$_3$Si$_2$Te$_6$ may be narrower than that of visible light of $1.64\sim3.19$ eV. We note the fitted thermal activation energy at $350\sim400$ K may be much smaller than the band gap.

\begin{figure}[t]
\includegraphics[scale=0.55]{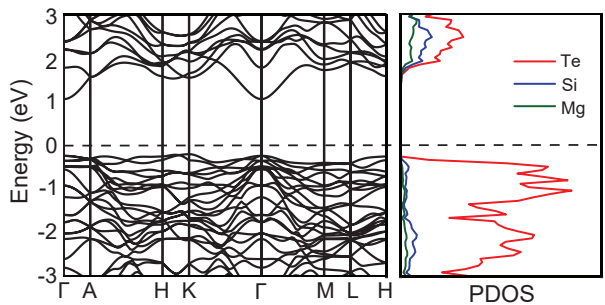}
\caption{Electronic structure calculation using DFT. The direct gap of Mg$_3$Si$_2$Te$_6$ is about 1.2 eV. PDOS is obtained by integrating density states of the energy bands, and the electrons close to the fermi surface are mainly provided by Te.}
\label{fig5}
\end{figure}

UV-vis absorption spectra were employed to study the electronic band gap of Mg$_3$Si$_2$Te$_6$. Five pieces of single crystals were measured separately. The absorption spectra are shown in Fig. \ref{fig4} (a), yielding a typical semiconducting state with a strong absorption in the range of $700\sim860$ nm, while the rest of the short wavelength data are removed because of relatively high noise. The band gap can be obtained by fitting the absorption spectra using the Tauc relation that is [$F$($R$$_\infty$)$h$$\nu$]$^{1/n}$ = $A$($h$$\nu$ - $E_g$), where $F$($R$$_\infty$) is the Kubelka-Mubelka-Munk function, $h$ is the Planck constant, $\nu$ is the frequency, $A$ is a prefactor, and $E_g$ is the band gap energy\cite{sarkar2017}. For a direct band gap semiconductor,  $n$ is 1/2; for an indirect band gap semiconductor, $n$ is 2. We plot the Tauc relations as a direct gap and indirect gap semiconductor in Figs. \ref{fig4} (b) and \ref{fig4} (c), respectively. The size of the band gap is extracted from the linear regression at the inflection point and the obtained $h\nu$-intercept value is taken as the band gap value\cite{tauc1966}. The measured absorption spectra could not distinguish the two models due to poor statistic. We fitted the linear parts of five sets of data by adopting the Tauc relation and $n=1/2$, resulting in the direct band gaps of 1.41, 1.43, 1.43, 1.37, and 1.31 eV. Thus, the direct gap is estimated to be 1.39(5) eV by averaging the fitted values.  Following the same procedure and adopting $n=2$, an indirect band gap of 0.6(2) eV is obtained. The difference in the calculated energy gaps between different data sets can be attributed to the inconsistency of thicknesses of the samples, which causes calculation errors by affecting the transmittance and reflectivity of the samples to ultraviolet light.

In order to distinguish the direct gap and indirect gap from the fittings of the UV-vis absorption spectra, DFT calculations were performed to investigate the electronic band structure. Figure \ref{fig5} displays the band structures along high symmetry directions in a Brillouin zone and partial density of states (PDOS) of Mg$_3$Si$_2$Te$_6$. The calculated data reveal a direct gap of $\sim$1.2 eV at the $\Gamma$ point, close to the fitting results from the UV-vis absorption spectra using the direct gap model. Thus, Mg$_3$Si$_2$Te$_6$ can be considered as a direct gap semiconductor with a gap size of $1.2\sim1.39$ eV. This value is larger than the thermal activation energy of 0.378 eV fitted from the resistivity between 350 and 400 K. The outcomes may attribute to the fact that the activation energy is fitted from resistance change at high temperature, where thermal fluctuations increase the electron hopping capability, and the gap seems to be narrowed. The direct gap size of Mg$_3$Si$_2$Te$_6$ is smaller than the energies of visible light. Through doping, temperature, and pressure, the band gap of Mg$_3$Si$_2$Te$_6$ may be further decreased, allowing Mg$_3$Si$_2$Te$_6$  one of the potential candidates for infrared photoelectric materials.

\section{SUMMARY}

In conclusion, we synthesized single crystals of Mg$_3$Si$_2$Te$_6$ by self-flux method, and characterized the structure, composition, resistivity, and electronic band gap. Mg$_3$Si$_2$Te$_6$ exhibits a quasi-2D structure constituted by Mg$_2$Si$_2$Te$_6$ layers (Mg1) that are linked by Mg2 atoms. The single crystals can be mechanically cleaved and are sensitive to air. By combining the UV-vis absorption spectra and DFT calculations, we demonstrate that Mg$_3$Si$_2$Te$_6$ is a direct band gap semiconductor with a gap size of $1.2\sim1.39$ eV.
This gap size is close to that of silicon, and may allow Mg$_3$Si$_2$Te$_6$ become one of the candidates for infrared photoelectric or other functional materials by tuning the energy level by rare earth ions doping.
Moreover, the direct gap characteristics may provide higher photon absorption efficiency compared with traditional indirect-gap Si-based semiconductors. The absorption capacity and conversion efficiency require further investigations.

\section{ACKNOWLEDGMENTS}

Work was supported by the National Natural Science Foundation of China (Grants No. 12174454, No. 11904414, No. 11904416, and No. 12104427), the Guangdong Basic and Applied Basic Research Foundation (Grant No. 2021B1515120015), and National Key Research and Development Program of China (Grant No. 2019YFA0705702).

\bibliography{ref}
\end{document}